\documentclass[aps,prd,twocolumn,superscriptaddress,showpacs]{revtex4}
\usepackage{epsfig,epsf}
\usepackage{amsmath}
\usepackage{amsthm}
\usepackage{amsfonts}
\usepackage{amssymb}
\usepackage{dsfont}
\usepackage{multirow}
\usepackage{appendix}
\usepackage{slashed}
\usepackage[active]{srcltx}
\usepackage{psfrag}

\begin{document}

\title{ Mass and decay constant of the newly observed exotic $X(5568)$ state }
\date{\today}
\author{S.~S.~Agaev}
\affiliation{Department of Physics, Kocaeli University, 41380 Izmit, Turkey}
\affiliation{Institute for Physical Problems, Baku State University, Az--1148 Baku,
Azerbaijan}
\author{K.~Azizi}
\affiliation{Department of Physics, Do\v{g}u\c{s} University, Acibadem-Kadik\"{o}y, 34722
Istanbul, Turkey}
\author{H.~Sundu}
\affiliation{Department of Physics, Kocaeli University, 41380 Izmit, Turkey}

\begin{abstract}
The mass and decay constant of the $X(5568)$ state newly observed
 by  D0 Collaboration are computed within the two-point
sum rule method using the diquark-antidiquark interpolating
current. In calculations, the vacuum condensates up to eight dimensions
are taken into account. The obtained result for the mass of
the $X(5568)$ state is in a nice agreement 
with the experimental data.

\end{abstract}

\maketitle

\section{Introduction}

During the last decade, due to a wide flow of experimental information
rushing from Belle, BABAR, BESIII, LHCb, CDF, D0 and some other
collaborations, investigation of exotic states, i.e. states that can not be
included into the quark-antiquark and three-quark bound schemes of the
traditional hadron spectroscopy, became one of the interesting and growing
fields in the hadron physics. The period of intensive experimental and
theoretical studies of exotic particles started from the discovery of the
charmonium-like resonance $X(3872)$ by Belle Collaboration \cite{Belle:2003}%
, confirmed later in some other experiments \cite%
{D0:2004,CDF:2004,Babar:2005}. The exotic states were produced and observed
in $B$ meson decays, in the $e^{+}e^{-}$ and $\overline{p}p$ annihilations,
in the double charmonium production processes, in the two-photon fusion and $%
pp$ collisions. Experimental investigations encompass measurements of the
masses and decay widths of these states, exploration of their spins,
parities and C-parities.  The charmonium-like  exotic states observed and studied
till now form the wide family  of XYZ particles.

These discoveries necessitated generation of new theoretical approaches to
interpret underlying quark-gluon structure of the exotic states, and
invention of methods for calculation of their properties. (see, the reviews
\cite%
{Swanson:2006st,Klempt:2007cp,Godfrey:2008nc,Voloshin:2007dx,Nielsen:2010,
Faccini:2012pj,Esposito:2014rxa,Chen:2016} and references therein).
Naturally, efforts were done to consider new charmonium-like resonances as
excitations of the ordinary $c\overline{c}$ charmonium and describe their
features applying updated quark-antiquark potentials. It should be noted
that some of new resonances really allow interpretation as the excited $c%
\overline{c}$ states. But the main part of the collected experimental data
can not be entered into this frame, and therefore for understanding of the
phenomenology of XYZ states various quark-gluon models were suggested.

One of the mostly employed models is the four-quark or tetraquark picture of
the exotic states. In accordance with this approach new charmonium-like
states are formed by two heavy and two light quarks. These quarks may
cluster into the colored diquark and antidiquark, which are organized in
such a way that to reproduce quantum numbers of the corresponding exotic
states \cite{Maiani:2004vq}. This tetraquark model is known as
diquark-antidiquark model of the exotic states. In the meson-molecule model
the exotic particle may appear as a bound state of two color-singlet mesons.
There are other models within the tetraquark approach Refs.\cite%
{Braaten:2013oba,Dubynskiy:2008mq}, as well as ones that exploit alternative
ideas (see, for example Ref.\ \cite{Braaten:2013boa}).

Recently, the D0 Collaboration reported the observation of a narrow
structure $X(5568)$ in the decay chain $X(5568) \to B_{s}^{0} \pi^{\pm}$, $%
B_{s}^{0} \to J/\psi \phi$, $J/\psi \to \mu^{+} \mu^{-}$, $\phi \to
K^{+}K^{-}$ \cite{D0:2016mwd} based on $p\bar{p}$ collision data at $\sqrt{s}%
=1.96\ \mathrm{TeV}$ collected at the Fermilab Tevatron collider. In order to distinguish it
from the "traditional" members of the X family of exotic resonances, in what
follows we will use for this state the notation $%
X_b(5568)$. As it was emphasized in Ref.\ \cite{D0:2016mwd} this is the
first observation of a hadronic state with four quarks of different flavors.
Namely, from the observed decay channel $X_b(5568) \to B_{s}^{0} \pi^{\pm}$
it is not difficult to conclude that the state $X_{b}(5568)$ consists of $%
b,\, s, \, u, \, d$ quarks. The assigned quantum numbers for the $X_{b}$
state are $J^{PC}=0^{++}$, its mass extracted from the experiment is equal to $%
m_{X_{b}}=5567.8 \pm 2.9 \mathrm{(stat)}^{+0.9}_{-1.9} \mathrm{(syst)}\,
\mathrm{MeV}$, and the decay width was estimated as $\Gamma=21.9 \pm 6.4
\mathrm{(stat)}^{+5.0}_{-2.5} \mathrm{(syst)}\, \mathrm{MeV}$. First
suggestions concerning the quark-antiquark organization of the new state
were made in Ref.\ \cite{D0:2016mwd}, as well. Thus, within the diquark-antidiquark
model the $X_b$ may be described as $[bu][\bar{d}\bar{s}]$, $[bd][\bar{s}%
\bar{u}]$, $[su][\bar{b}\bar{d}]$ or $[sd][\bar{b}\bar{u}]$ bound state.
Alternatively, it may be considered as a molecule composed of B and K mesons.

In the present work for the $X_b$ we adopt $[su][\bar{b}\bar{d}]$ diquark
model, and calculate for the first time its mass and decay constant using
the QCD two-point sum rule.

This article is organized in the following manner. In section \ref{sec:Mass}%
, we calculate the mass and decay constant of the $X_{b}$ state employing
two-point QCD sum rule approach including into analysis the vacuum
condensates up to eighth dimension. Our numerical results are presented in
Section \ref{sec:Num} and compared with the
experimental data of the D0 Collaboration. This section contains also our concluding
remarks. The explicit expressions of the
spectral density required for calculation of the mass and decay constant are
written down in Appendix A.


\section{The sum rules for the mass and decay constant}

\label{sec:Mass}

To calculate the mass and decay constant of the $X_{b}$ state in the framework
of QCD sum rules, we start from the two-point correlation function
\begin{equation}
\Pi (p)=i\int d^{4}xe^{ip\cdot x}\langle 0|\mathcal{T}\{J^{X_{b}}(x)J^{X_{b}%
\dag }(0)\}|0\rangle ,  \label{eq:CorrF1}
\end{equation}%
where $J^{X_{b}}(x)$ is the interpolating current with required quantum
numbers. We consider $X_{b}(5568)$ state as a particle with the quantum
numbers $J^{PC}=0^{++}$. Then in the diquark model the current $J^{X_{b}}(x)$
is defined by the following expression
\begin{equation}
J^{X_{b}}(x)=\varepsilon ^{ijk}\varepsilon ^{imn}\left[ s^{j}(x)C\gamma
_{\mu }u^{k}(x)\right] \left[ \overline{b}^{m}(x)\gamma ^{\mu }C\overline{d}%
^{n}(x)\right] .  \label{eq:CDiq}
\end{equation}
In Eq.\ (\ref{eq:CDiq}) $i,\ j,\ k,m,\ n$ are color indexes and $C$ is the
charge conjugation matrix.

In order to derive QCD sum rule expression we calculate the correlation
function in terms of the physical degrees of freedom. Performing integral
over $x$ in Eq.\ (\ref{eq:CorrF1}), we get
\begin{equation*}
\Pi ^{\mathrm{Phys}}(p)=\frac{\langle 0|J^{X_{b}}|X_{b}(p)\rangle \langle
X_{b}(p)|J^{X_{b}\dag }|0\rangle }{m_{X_{b}}^{2}-p^{2}}+...
\end{equation*}%
where $m_{X_{b}}$ is the mass of the $X_{b}(5568)$ state, and dots stand for
contributions of the higher resonances and continuum states. We define the
decay constant $f_{X_{b}}$ through the matrix element%
\begin{equation*}
\langle 0|J^{X_{b}}|X_{b}(p)\rangle =f_{X_{b}}m_{X_{b}}.
\end{equation*}
Then in terms of $m_{X_{b}}$ and $f_{X_{b}}$ the correlation function can be
written in the form
\begin{equation}
\Pi ^{\mathrm{Phys}}(p)=\frac{m_{X_{b}}^{2}f_{X_{b}}^{2}}{m_{X_{b}}^{2}-p^{2}%
}+\ldots  \label{eq:CorM}
\end{equation}%
The Borel transformation applied to Eq.\ (\ref{eq:CorM}) yields%
\begin{equation}
\mathcal{B}_{p^{2}}\Pi ^{\mathrm{Phys}%
}(p)=m_{X_{b}}^{2}f_{Xb}^{2}e^{-m_{X_{b}}^{2}/M^{2}}+\ldots
\label{eq:CorBor}
\end{equation}

The theoretical expression for the same function, $\Pi ^{\mathrm{QCD}}(p)$,
has to be determined employing of the quark-gluon degrees of freedom. To
this end, we contract the heavy and light quark fields, and for the
correlation function $\Pi ^{\mathrm{QCD}}(p)$ find:
\begin{eqnarray}
&&\Pi ^{\mathrm{QCD}}(p)=i\int d^{4}xe^{ipx}\varepsilon ^{ijk}\varepsilon
^{imn}\varepsilon ^{i^{\prime }j^{\prime }k^{\prime }}\varepsilon
^{i^{\prime }m^{\prime }n^{\prime }}  \notag \\
&&\times \mathrm{Tr}\left[ \gamma _{\mu }\widetilde{S}_{d}^{n^{\prime
}n}(-x)\gamma _{\nu }S_{b}^{m^{\prime }m}(-x)\right]  \notag \\
&&\times\mathrm{Tr}\left[ \gamma ^{\nu }\widetilde{S}_{s}^{jj^{\prime
}}(x)\gamma ^{\mu }S_{u}^{kk^{\prime }}(x)\right] .  \label{eq:CorrF2}
\end{eqnarray}
In Eq.\ (\ref{eq:CorrF2}) we introduce the notation
\begin{equation*}
\widetilde{S}_{q}^{ij}(x)=CS_{q}^{ijT}(x)C,
\end{equation*}%
with $S_{q}^{ij}(x)$ and $S_{b}^{ij}(x)$ being the light ($q\equiv u,\ d\ $%
and $s$) and heavy quark propagators, respectively. We choose the light
quark propagator $S_{q}^{ij}(x)$ in the $x$-space in the form%
\begin{eqnarray}
&&S_{q}^{ij}(x)=i\delta _{ij}\frac{\slashed x}{2\pi ^{2}x^{4}}-\delta _{ij}%
\frac{m_{q}}{4\pi ^{2}x^{2}}-\delta _{ij}\frac{\langle \overline{q}q\rangle
}{12}  \notag \\
&&+i\delta _{ij}\frac{\slashed xm_{q}\langle \overline{q}q\rangle }{48}%
-\delta _{ij}\frac{x^{2}}{192}\langle \overline{q}g\sigma Gq\rangle +i\delta
_{ij}\frac{x^{2}\slashed xm_{q}}{1152}\langle \overline{q}g\sigma Gq\rangle
\notag \\
&&-i\frac{gG_{ij}^{\alpha \beta }}{32\pi ^{2}x^{2}}\left[ \slashed x{\sigma
_{\alpha \beta }+\sigma _{\alpha \beta }}\slashed x\right] -i\delta _{ij}%
\frac{x^{2}\slashed xg^{2}\langle \overline{q}q\rangle ^{2}}{7776}  \notag \\
&& -\delta _{ij}\frac{x^{4}\langle \overline{q}q\rangle \langle
g^{2}GG\rangle }{27648}+ \ldots  \label{eq:qprop}
\end{eqnarray}%
For the heavy quark propagator $S_{b}^{ij}(x)$ we employ the expression \cite%
{Reinders:1984sr}
\begin{eqnarray}
&&S_{b}^{ij}(x)=i\int \frac{d^{4}k}{(2\pi )^{4}}e^{-ikx}\left[ \frac{\delta
_{ij}\left( {\slashed k}+m_{b}\right) }{k^{2}-m_{b}^{2}}\right.  \notag \\
&&-\frac{gG_{ij}^{\alpha \beta }}{4}\frac{\sigma _{\alpha \beta }\left( {%
\slashed k}+m_{b}\right) +\left( {\slashed k}+m_{b}\right) \sigma _{\alpha
\beta }}{(k^{2}-m_{b}^{2})^{2}}  \notag \\
&&\left. +\frac{g^{2}}{12}G_{\alpha \beta }^{A}G^{A\alpha \beta }\delta
_{ij}m_{b}\frac{k^{2}+m_{b}{\slashed k}}{(k^{2}-m_{b}^{2})^{4}}+\ldots %
\right] .  \label{eq:Qprop}
\end{eqnarray}%
In Eqs.\ (\ref{eq:qprop}) and (\ref{eq:Qprop}) the short-hand notation
\begin{equation*}
G_{ij}^{\alpha \beta }\equiv G_{A}^{\alpha \beta
}t_{ij}^{A},\,\,\,\,A=1,\,2\,\ldots 8,
\end{equation*}%
is used, where $i,\,j$ are color indexes, and $t^{A}=\lambda ^{A}/2$ with $%
\lambda ^{A}$ being the standard Gell-Mann matrices. The first term in Eq.\ (%
\ref{eq:Qprop}) is the free (perturbative) massive quark propagator, next
ones are nonperturbative gluon corrections. In the nonperturbative terms the
gluon field strength tensor $G_{\alpha \beta }^{A}\equiv G_{\alpha \beta
}^{A}(0)$ is fixed at $x=0.$

In general, the QCD sum rule expressions are derived after fixing the same
Lorentz structures in the both phenomenological and theoretical expressions
of the correlation function. In the case under consideration this structure
is trivial and $\sim \mathrm{I}$. Then there is only one invariant function $%
\Pi ^{\mathrm{QCD}}(p^{2})$, which can be written down as the dispersion
integral
\begin{equation}
\Pi ^{\mathrm{QCD}}(p^{2})=\int_{(m_{b}+m_{s})^{2}}^{\infty }\frac{\rho ^{%
\mathrm{QCD}}(s)}{s-p^{2}}+...,
\end{equation}%
where $\rho ^{\mathrm{QCD}}(s)$ is the corresponding spectral density.

The problem posed in this section is calculation of the spectral density $%
\rho ^{\mathrm{QCD}}(s)$ necessary for the mass and decay constant analysis.
In the present work we include into the sum rule calculations the quark,
gluon and mixed vacuum condensates up to and including ones with the
dimension 8. For computation of the components of the spectral density we
use the technique, essential steps of which and main formulas for their
realization were provided in Ref. \cite{Agaev:2016dev}. This computational
scheme includes the following stages: we apply the integral transformation
for the terms $\sim 1/(x^{2})^{n}$ coming from the light quark propagators,
when necessary replace $x_{\mu }$ by $-i\partial /\partial q_{\mu }$, and
then calculate the obtained $x$ integral. The Dirac delta function appeared
in a result of such integration allows us to remove one of the momentum
integrals. In order to carry out the remaining integration we use the
Feynman parametrization rearranging denominators obtained after this
operation, and derive the final expressions applying the well known formulas
\cite{Agaev:2016dev}. The imaginary part of the correlation function can now
be extracted by applying in the $D\rightarrow 4$ limit the replacement
\begin{equation}
\Gamma \left( \frac{D}{2}-n\right) \left( -\frac{1}{L}\right) ^{\frac{D}{2}%
-n}\rightarrow \frac{(-1)^{n-1}}{(n-2)!}(-L)^{n-2}\ln (-L).
\end{equation}%
As a result, we get the imaginary part of the correlation function, and
hence the components of the spectral density as the integrals over the
Feynman parameter $z$. The expressions derived for $\rho ^{\mathrm{QCD}}(s)$
in accordance with these recipes are collected in Appendix A.

Applying the Borel transformation on the variable $p^{2}$ to the invariant
amplitude $\Pi ^{\mathrm{QCD}}(p^{2})$, equating the obtained expression
with the relevant part of $\mathcal{B}_{p^{2}}\Pi ^{\mathrm{Phys}}(p)$, and
subtracting the continuum contribution, we finally obtain the required sum
rule. Thus, the mass of the $X_{b}$ state can be evaluated from the sum rule
\begin{equation}
m_{X_{b}}^{2}=\frac{\int_{(m_{b}+m_{s})^{2}}^{s_{0}}dss
\rho ^{\mathrm{QCD}}(s)e^{-s/M^{2}}}{\int_{(m_{b}+m_{s})^{2}}^{s_{0}}ds\rho
^{\mathrm{QCD}}(s)e^{-s/M^{2}}},  \label{eq:srmass}
\end{equation}%
whereas to extract the numerical value of the decay constant $f_{X_{b}}$ we
employ the formula
\begin{equation}
f_{X_{b}}^{2}m_{X_{b}}^{2}e^{-m_{X_{b}}^{2}/M^{2}}=%
\int_{(m_{b}+m_{s})^{2}}^{s_{0}}ds\rho ^{\mathrm{QCD}}(s)e^{-s/M^{2}}.
\label{eq:srcoupling}
\end{equation}%
The Eqs.\ (\ref{eq:srmass}) and (\ref{eq:srcoupling}) are the sum rules
required for evaluating of the $X_{b}$ state's mass and decay constant,
respectively.



\section{Numerical results and conclusions}

\label{sec:Num}
\begin{table}[tbp]
\begin{tabular}{|c|c|}
\hline\hline
Parameters & Values \\ \hline\hline
$m_{b}$ & $(4.18\pm0.03)~\mathrm{GeV}$ \\
$m_{s} $ & $(95 \pm 5)~\mathrm{MeV} $ \\
$\langle \bar{q}q \rangle $ & $(-0.24\pm 0.01)^3 $ $\mathrm{GeV}^3$ \\
$\langle \bar{s}s \rangle $ & $0.8\ \langle \bar{q}q \rangle$ \\
$\langle\frac{\alpha_sG^2}{\pi}\rangle $ & $(0.012\pm0.004)$ $~\mathrm{GeV}%
^4 $ \\
$ m_{0}^2 $ & $(0.8\pm0.1)$ $\mathrm{GeV}^2$ \\
$\langle \overline{q}g\sigma Gq\rangle $ & $m_{0}^2\langle \bar{q}q \rangle$  \\ \hline\hline
\end{tabular}%
\caption{Input parameters used in calculations}
\label{tab:Param}
\end{table}

\begin{figure}[tbp]
\centerline{
\begin{picture}(210,170)(0,0)
\put(-10,0){\epsfxsize8.2cm\epsffile{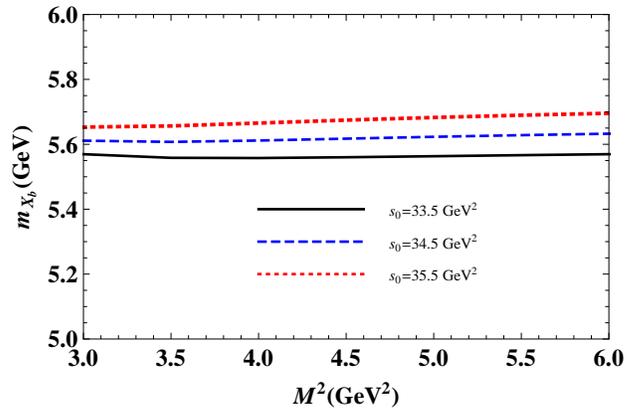}}
\end{picture}
}
\caption{The mass $m_{X_{b}}$ as a function of the Borel parameter $M^2$ for
different values of $s_0$.}
\label{fig:Xbmass}
\end{figure}

\begin{figure}
\centerline{
\begin{picture}(200,170)(0,0)
\put(-10,20){\epsfxsize8.2cm\epsffile{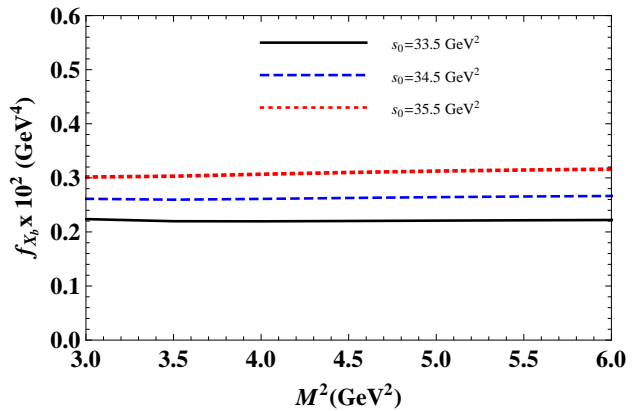}}
\end{picture}}
\caption{The decay constant $f_{X_{b}}$ vs Borel parameter $M^2$. The values
of the parameter $s_0$ are shown in the figure.}
\label{fig:Xbcoup}
\end{figure}

The QCD sum rules expressions for the mass and decay constant of the $X_{b}$
contain various parameters that should be fixed in accordance with the
standard procedures. Thus, for numerical computation of the $X_{b}$ state's mass
and decay constant we need values of the quark, gluon and mixed condensates.
In addition to that, QCD sum rules depend on the $b$ and $s$ quark masses.
The values of these parameters can be found in Table \ref{tab:Param}.

Sum rules calculations require fixing of the threshold parameter $s_{0}$ and
a region within of which it may be varied. For $s_{0}$ we employ
\begin{equation}
33.5\,\,\mathrm{GeV}^2 \leq s_{0}\leq 35.5 \,\,\mathrm{GeV}^2.
\end{equation}
We also find the range $3\ \mathrm{GeV}^2<M^2<6\ \mathrm{GeV}^2$
as a reliable one for varying the Borel parameter, where the
effects of the higher resonances and continuum states, and
contributions of the higher dimensional condensates meet all
requirements of QCD sum rules calculations. Additionally, in this
interval the dependence of the mass and decay constant on $M^2$ is
stable, and we may expect that the sum rules give the correct
results. By varying the parameters $M^2$ and $s_{0}$ within the
allowed ranges, as well as taking into account ambiguities arising
from other input parameters we estimate uncertainties of the whole
calculations. The results for the mass $m_{X_{b}}$ and decay
constant $f_{X_{b}}$ are depicted as the functions of the Borel
parameter in Figs.~\ref{fig:Xbmass} and \ref{fig:Xbcoup},
respectively. The sensitivity of the obtained predictions to the
choice of $s_0$ are also seen in these figures, where three
different values for $s_0$ are employed. Our prediction for the
mass $m_{X_{b}}$ is:
\begin{equation}
m_{X_{b}}=(5584\pm 137)\, \mathrm{MeV}.
\end{equation}
For the decay constant we get:
\begin{equation}
f_{X_b}=(0.24\pm 0.02)\times 10^{-2}\,\, \mathrm{GeV}^{4}.
\end{equation}
As is seen our prediction for the mass of the $X_{b}(5568)$ state agrees with
experimental data of the D0 Collaboration.

In this paper we have studied the new exotic resonance state with the mass
$5568\, \mathrm{MeV}$ and quantum numbers $J^{PC}=0^{++}$, that was observed
recently by D0 Collaboration utilizing the collected data of $p \bar{p}$ collision.
We have adopted for this state a label $X_b(5568)$ because it is composed of four
different quark flavors and differs from the usual  charmonium-like members of the
X family. We have also accepted  the diquark-antidiquark model $[su][\bar{b}\bar{d}]$
for  the $X_b(5568)$ state and computed its mass and decay constant employing QCD
two-point sum rule method. Our prediction for the mass $m_{X_{b}}$ is in agreement
with the finding of D0 Collaboration. Results of our explorations of the $X_b(5568)$
state obtained by applying other diquark-antidiquark structures and interpolating currents as well as calculation of its decay width, which can be
crucial in making decision between various models, will be published elsewhere.

\section*{ACKNOWLEDGEMENTS}

The work of S.~S.~A. was supported by the TUBITAK grant 2221-"Fellowship
Program For Visiting Scientists and Scientists on Sabbatical Leave". This
work was also supported in part by TUBITAK under the grant no: 115F183.

\appendix*

\section{A}

\renewcommand{\theequation}{\Alph{section}.\arabic{equation}}

\label{sec:App} In this appendix we have collected the results of our
calculations of the spectral density
\begin{equation}
\rho ^{\mathrm{QCD}}(s)=\rho ^{\mathrm{pert}}(s)+\sum_{k=3}^{k=8}\rho
_{k}(s),  \label{eq:A1}
\end{equation}%
used for evaluation of the $X_{b}$ meson mass $m_{X_{b}}$ and its decay
constant $f_{X_{b}}$ from the QCD sum rule. In Eq.\ (\ref{eq:A1}) by $\rho
_{k}(s)$ we denote the nonperturbative contributions to $\rho ^{\mathrm{QCD}%
}(s)$. In calculations we have neglected the masses of the $u$ and $d$
quarks and taken into account terms $\sim m_s$. The explicit expressions for
$\rho ^{\mathrm{pert}}(s)$ and $\rho _{k}(s)$ are presented below as the
integrals over the Feynman parameter $z$:
\begin{widetext}
\begin{eqnarray}
&&\rho ^{\mathrm{pert}}(s)=\frac{1}{1536\pi ^{6}}\int\limits_{0}^{a}\frac{%
dzz^{4}}{(z-1)^{3}}\left[ m_{b}^{2}+s(z-1)\right]^3 \left[ m_{b}^{2}+3s(z-1)%
\right] , \notag \\
&&\rho _{\mathrm{3}}(s)=\frac{1}{32\pi ^{4}}\int\limits_{0}^{a}\frac{dzz^{2}}{%
(z-1)^{2}}\left[ m_{b}^{2}+s(z-1)\right] \left\{ \langle \overline{d}%
d\rangle m_{b}\left[ m_{b}^{2}+s(z-1)\right] +2m_{s}(\langle \overline{s}%
s\rangle -\langle \overline{u}u\rangle )\left[ m_{b}^{2}+2s(z-1)\right]
(z-1)\right\} , \notag \\
&&\rho _{\mathrm{4}}(s)=\frac{1}{2304\pi ^{4}}\langle \alpha _{s}\frac{G^{2}}{%
\pi }\rangle \int\limits_{0}^{a}\frac{dzz^{2}}{(z-1)^{3}}\left\{ m_{b}^{4}%
\left[ z(8z-15)+9\right] +3m_{b}^{2}s(z-1)\left[ z(7z-15)+9\right]
+6s^{2}(z-1)^{3}(2z-3)\right\} , \notag \\
&&\rho _{\mathrm{5}}(s)=\frac{m_{0}^{2}}{192\pi ^{4}}\int\limits_{0}^{a}\frac{%
dzz}{(1-z)}\left\{ 3m_{b}\langle \overline{d}d\rangle \left[ m_{b}^{2}+s(z-1)%
\right] +m_{s}(z-1)(2\langle \overline{s}s\rangle -3\langle \overline{u}%
u\rangle )\left[ 2m_{b}^{2}+3s(z-1)\right] \right\} , \notag \\
&&\rho _{\mathrm{6}}(s)=\frac{1}{324\pi ^{4}}\int\limits_{0}^{a}dzzg^{2}(%
\langle \overline{u}u\rangle ^{2}+\langle \overline{d}d\rangle ^{2}+\langle
\overline{s}s\rangle ^{2})\left[ 2m_{b}^{2}+3s(z-1)\right] , \notag \\
&&\rho _{\mathrm{7}}(s)=\frac{1}{576\pi ^{2}}\langle \alpha _{s}\frac{G^{2}}{%
\pi }\rangle \int\limits_{0}^{a}\frac{dz}{(1-z)}\left\{ 2m_{b}\langle
\overline{d}d\rangle (5z-2)+m_{s}(z-1)\left[ 3\langle \overline{s}s\rangle
+4\langle \overline{u}u\rangle (4z-1)\right] \right\}, \notag \\
&&\rho_8(s)=-\frac{11}{9216 \pi^2}\langle \alpha_s
\frac{G^2}{\pi}\rangle^2\frac{(m_b^2-s)^2}{s^2}-m_0^2\langle
\bar{s}s\rangle\langle\bar{u}u\rangle \int\limits_{0}^{a} dz\frac{z-1}{6\pi^2},
\end{eqnarray}%
\end{widetext}
where $a=(s-m_{b}^{2})/s.$

\end{document}